\documentclass[preprint,showpacs,preprintnumbers,amsmath,amssymb,nofootinbib]{revtex4}
\usepackage[dvipdfmx]{graphicx}
\usepackage{amsmath,amssymb}
\usepackage{bm}% bold math
\usepackage{color} % For blue in-text comments and additions

\def\lesssim{\ \raise.3ex\hbox{$<$}\kern-0.8em\lower.7ex\hbox{$\sim$}\ }
\def\gesim{\ \raise.3ex\hbox{$>$}\kern-0.8em\lower.7ex\hbox{$\sim$}\ }
\def\bm#1{\mathbf{#1}}

\begin{document}

%=================================================================
% Full title of the paper (Capitalized)
\title{BCS-BEC Crossover and Pairing Fluctuations in a Two Band Superfluid/Superconductor: A {\it T} Matrix Approach}

% Author Orchid ID: enter ID or remove command
%\newcommand{\orcidauthorA}{0000-0001-5247-7116} % Add \orcidA{} behind the author's name
%\newcommand{\orcidauthorB}{0000-0001-8295-805X} % Add \orcidB{} behind the author's name
%\newcommand{\orcidauthorC}{0000-0002-4914-4975}

% Authors, for the paper (add full first names)
\author{Hiroyuki Tajima$^{1,2}$, Andrea Perali$^{3}$, and Pierbiagio Pieri$^{4,5}$}

% Authors, for metadata in PDF
%\AuthorNames{Firstname Lastname, Firstname Lastname and Firstname Lastname}

% Affiliations / Addresses (Add [1] after \address if there is only one affiliation.)
\affiliation{%
$^{1}$Department of Mathematics and Physics, Kochi University, Kochi 780-8520, Japan; htajima@kochi-u.ac.jp\\
$^{2}$Quantum Hadron Physics Laboratory, RIKEN Nishina Center, Wako, Saitama, 351-0198, Japan\\
$^{3}$School of Pharmacy, Physics Unit, Universit\`{a} di Camerino, 62032 Camerino (MC), Italy\\
$^{4}$School of Science and Technology, Physics Division, Universit\`{a} di Camerino, 62032 Camerino (MC), Italy\\
$^{5}$INFN, Sezione di Perugia, 06123 Perugia (PG), Italy}

% Contact information of the corresponding author
%\corres{Correspondence: }

% Current address and/or shared authorship
%\firstnote{Current address: Affiliation 3} 
%\secondnote{These authors contributed equally to this work.}
% The commands \thirdnote{} till \eighthnote{} are available for further notes

%\simplesumm{} % Simple summary

%\conference{} % An extended version of a conference paper

% Abstract (Do not insert blank lines, i.e. \\) 
\begin{abstract}
We investigate pairing fluctuation effects in a two band fermionic system, where a shallow band in the Bardeen--Cooper--Schrieffer--Bose--Einstein condensation (BCS-BEC) crossover regime is coupled with a weakly interacting deep band.
Within a diagrammatic $T$ matrix approach, we~report how thermodynamic quantities such as the critical temperature, chemical potential, {and~momentum distributions} undergo the crossover from the BCS to BEC regime by tuning the intraband coupling in the shallow band.
We also generalize the definition of Tan's contact to a two band system and report the two contacts for different pair-exchange couplings.
The present results are compared with those obtained by the simpler Nozi\`eres--Schmitt--Rink approximation. We confirm a pronounced enhancement of the critical temperature due to the multiband configuration, as well as to the pair-exchange coupling.
\end{abstract}
\maketitle
% Keywords
%\keyword{multi-band superconductivity; BCS-BEC crossover; ultracold Fermi gases }

% The fields PACS, MSC, and JEL may be left empty or commented out if not applicable
%\PACS{J0101}
%\MSC{}
%\JEL{}

%%%%%%%%%%%%%%%%%%%%%%%%%%%%%%%%%%%%%%%%%%
% Only for the journal Diversity
%\LSID{\url{http://}}

%%%%%%%%%%%%%%%%%%%%%%%%%%%%%%%%%%%%%%%%%%
% Only for the journal Applied Sciences:
%\featuredapplication{Authors are encouraged to provide a concise description of the specific application or a potential application of the work. This section is not mandatory.}
%%%%%%%%%%%%%%%%%%%%%%%%%%%%%%%%%%%%%%%%%%

%%%%%%%%%%%%%%%%%%%%%%%%%%%%%%%%%%%%%%%%%%
% Only for the journal Data:
%\dataset{DOI number or link to the deposited data set in cases where the data set is published or set to be published separately. If the data set is submitted and will be published as a supplement to this paper in the journal Data, this field will be filled by the editors of the journal. In this case, please make sure to submit the data set as a supplement when entering your manuscript into our manuscript editorial system.}

%\datasetlicense{license under which the data set is made available (CC0, CC-BY, CC-BY-SA, CC-BY-NC, etc.)}

%%%%%%%%%%%%%%%%%%%%%%%%%%%%%%%%%%%%%%%%%%
% Only for the journal Toxins
%\keycontribution{The breakthroughs or highlights of the manuscript. Authors can write one or two sentences to describe the most important part of the paper.}

%\setcounter{secnumdepth}{4}
%%%%%%%%%%%%%%%%%%%%%%%%%%%%%%%%%%%%%%%%%%
%%%%%%%%%%%%%%%%%%%%%%%%%%%%%%%%%%%%%%%%%%

%%%%%%%%%%%%%%%%%%%%%%%%%%%%%%%%%%%%%%%%%%
\section{Introduction}
Superconductivity and superfluidity play a central role in modern physics. The first microscopic theory for superconductivity, namely the Bardeen--Cooper--Schrieffer (BCS) theory~\cite{BCS}, has successfully been applied to various quantum many body systems.
On the other hand, some of them, in particular so-called unconventional superconductors, involve nontrivial effects beyond the BCS theory due to their complex band structures.
In this regard, a theoretical model for multi-band superconductors was developed by Suhl, Matthias, and Walker in 1959~\cite{SMW}.
This pioneering work is the starting point for the recent studies of multi-component superconductors~\cite{Tanaka,Milosevic}.
In condensed matter systems,
such superconductors have experimentally been discovered, e.g., in 
MgB$_2$~\cite{Akimitsu} and iron based compounds~\cite{Kamihara}.
From the theoretical viewpoint, a variety of non-trivial phenomena has been proposed (e.g., topological phase soliton~\cite{Tanaka2,Yerin1}, odd-frequency pairing~\cite{Black}, multiple Leggett mode~\cite{Ota}, hidden criticality~\cite{Milosevic2}, stable Sarma phase~\cite{He1}, and~Lifshitz transitions with resonance effects and amplification of the critical temperature \cite{Valletta,Mazziotti}).
\par
As another direction toward the unconventional properties of superconductors,
the crossover from the weakly coupled BCS state to the Bose--Einstein condensate (BEC) of a tightly bound molecule with increasing the two body attractive interaction has extensively been discussed~\cite{Eagles,Leggett}.
In fact, this crossover has been realized in ultracold Fermi atomic gases~\cite{Regal,Zwierlein}, and currently, various properties and fluctuation phenomena of this system have been investigated theoretically and experimentally~\cite{Giorgini,Bloch,Strinati}.
Interestingly, recent experiments of FeSe multi-band superconductors also indicated that this electron system is in the BCS-BEC crossover regime~\cite{Kasahara1,Kasahara2,Rinott}.
\par
Combining these two aspects of unconventional superconducting/superfluid properties
is a promising route toward room temperature superconductivity~\cite{Innocenti,Salasnich}.
One striking feature predicted in the multi-band BCS-BEC crossover is the screening of pairing fluctuations~\cite{Salasnich,Tajima1}.
Since pairing fluctuations are known to lower generally the superconducting/superfluid critical temperature,
this~screening effect is expected to enhance the critical temperature compared to the single band counterpart.
Indeed, a missing pseudogap in the multi-band FeSe superconductors in the BCS-BEC crossover has been reported in a recent experiment~\cite{Hanaguri}.
Since single particle spectra in a single band system in the BCS-BEC crossover regime exhibit the pseudogap originating from strong pairing fluctuations~\cite{Perali,Tsuchiya,Palestini,Marsiglio}, this experimental finding supports the screening of pairing fluctuations due to the multi-band nature of the FeSe superconductor.
In addition, a recent torque magnetometry measurement also supports the absence of strong pairing fluctuations in this system~\cite{Takahashi}.
In contrast, other experimental works have reported superconducting fluctuation effects in FeSe multi-band superconductors~\cite{Kasahara2,Gati,Lubashevsky}.
In this sense, we have to establish a whole picture of this system to understand where multi-band FeSe superconductors locate in the BCS-BEC crossover regime.
In addition, since the realization of the multi-band BCS-BEC crossover has been anticipated in Yb Fermi atomic gases,
a systematic study is required to obtain a unified understanding of these unconventional strongly coupled systems~\cite{Zhang,Pagano,Hofer,He3,He2,Mondal}.
{Such a study could promote further interdisciplinary investigations of the BCS-BEC crossover, such as in nuclear systems and neutron stars~\cite{Strinati,Sedrakian,Ohashi-2019}.}
\par
In this work, we theoretically investigate the multi-band BCS-BEC crossover and the effects of pairing fluctuations within the framework of the many body $T$ matrix approximation, extending our previous work \cite{Tajima1}, which was based on the Nozi\`eres--Schmitt--Rink (NSR) approximation~\cite{NSR}.
{While this extension is crucial to access dynamical quantities such as spectral functions,
one has to examine also how higher fluctuation effects beyond the NSR scheme appear in thermodynamic quantities. Specifically, as first pointed out by Serene~\cite{Serene},
the many body $T$ matrix approximation sums all repeated scatterings of an electron by independent pair fluctuations, while omitting vertex corrections and interactions between fluctuations. The NSR approximation, as well, omits the latter, but it also misses the repeated scatterings by independent fluctuations.
}
Considering a two band configuration, where a strongly interacting shallow band is coupled with a weakly interacting deep band through the pair-exchange couplings~\cite{Wolf}, we~show the evolution of the critical temperature, {momentum distribution functions}, chemical potential, and~occupation number densities in each band.
{This specific band configuration is physically motivated by FeSe superconductors, which are believed to locate in the BCS-BEC crossover regime with a strongly interacting shallow band~\cite{Lubashevsky} and for which the BCS-BEC crossover can be tuned by chemical doping~\cite{Rinott}.
Therefore, we systematically investigate thermodynamic quantities in the two band model by changing the interaction in the shallow band from weak to strong.}
%By taking the analytic continuation of the Matsubara Green's function, we address the single-particle density of states to see effects of pairing fluctuations.
In this paper, we set $\hbar=k_{\rm B}=1$, and the volume is taken to be unity.
%The order of the section titles is: Introduction, Materials and Methods, Results, Discussion, Conclusions for these journals: aerospace,algorithms,antibodies,antioxidants,atmosphere,axioms,biomedicines,carbon,crystals,designs,diagnostics,environments,fermentation,fluids,forests,fractalfract,informatics,information,inventions,jfmk,jrfm,lubricants,neonatalscreening,neuroglia,particles,pharmaceutics,polymers,processes,technologies,viruses,vision

%%%%%%%%%%%%%%%%%%%%%%%%%%%%%%%%%%%%%%%%%%
\section{Hamiltonian}
%\begin{figure}[t]
%\centering
%\includegraphics[width=5 cm]{fig1.eps}
%\caption{Schematic picture of the band structure for the fermionic system considered in this work: two parabolic bands are shifted in energy by $E_0$ to form a shallow and deep bands with a common chemical potential.}
%\label{fig1}
%\end{figure} 

We start from the two band fermion model~\cite{Tajima1,Iskin,Yerin2} 
%(see also Figure \ref{fig1})
described by the Hamiltonian:
\begin{equation}
\label{eq1}
H=\sum_{\bm{k},\sigma,i}\xi_{\bm{k},i}c_{\bm{k},\sigma,i}^{\dag}c_{\bm{k},\sigma,i} +\sum_{i,j}U_{ij}\sum_{\bm{Q}}b_{\bm{Q},i}^{\dag}b_{\bm{Q},j}.
\end{equation}
where $c_{\bm{k},\sigma,i}$ is an annihilation operator of a fermion with momentum $\bm{k}$ and spin $\sigma=\uparrow,\downarrow$ in the $i$ band ($i=1,2$).
$\xi_{\bm{k},i}=\varepsilon_{\bm{k},i}-\mu$ is the kinetic energy measured from the chemical potential $\mu$.
We assume parabolic band structures with a common effective mass $m_1=m_2=m$ in the momentum space as $\varepsilon_{\bm{k},i}=k^2/(2m_i)+E_0\delta_{i,2}$ where $E_0$ is the energy separation between the two bands.
The second term in Equation (\ref{eq1}) is the interaction term with coupling constants $U_{ij}$ and the pair annihilation operator $b_{\bm{Q},i}=\sum_{\bm{k}}^{k_0}c_{-\bm{k+Q}/2,\downarrow,i}c_{\bm{k+Q}/2,\uparrow,i}$ in the $i$ band with center of  mass momentum $\bm{Q}$. $k_0$ is a momentum cutoff, taken to be $k_0= 100 k_{\rm F,t}$ where $k_{\rm F,t}=(3\pi^2 n)^{1/3}$ is the Fermi momentum of the total system with the total number density $n$.
The intraband couplings $U_{jj}$ can be characterized in terms of the intraband scattering lengths $a_{jj}$ as:
\begin{eqnarray}
\label{eq2}
\frac{m}{4\pi a_{jj}}=\frac{1}{U_{jj}}+\sum_{\bm{k}}^{k_0}\frac{m}{k^2}.
\end{eqnarray} 

In this work, we define the band Fermi momentum $k_{{\rm F},i}=(3\pi^2 n_{0,i})^{1/3}$ where $n_{0,i}$ is the number density of each band in the absence of interactions at zero temperature, noting that $n=n_{1,0}+n_{2,0}$.
Then, we fix $(k_{\rm F,1}a_{11})^{-1}=-4$ while $(k_{\rm F,2}a_{22})^{-1}$ is tuned from the weak coupling BCS to the strong coupling BEC regime.
For convenience, we also introduce a dimensionless pair-exchange coupling 
$\lambda_{12}=U_{12}(k_0/k_{\rm F,t})^2n/E_{\rm F,t}$ (where $E_{\rm F,t}=k_{\rm F,t}^2/(2m)$) and $U_{21}=U_{12}$. 
To realize the situation where the shallow band is close to the Lifshitz transition~\cite{Wolf},
we take $\zeta \equiv E_0/E_{\rm F,1}=0.6$, corresponding to $n_{1,0}=[1+(1-\zeta)^{\frac{3}{2}}]^{-1}n\simeq 0.798n$. 
%%%%%%%%%%%%%%%%%%%%%%%%%%%%%%%%%%%%%%%%%%
\section{\textit{T} Matrix Approximation for Two Band Systems}
\begin{figure}[t]
\centering
\includegraphics[width=6 cm]{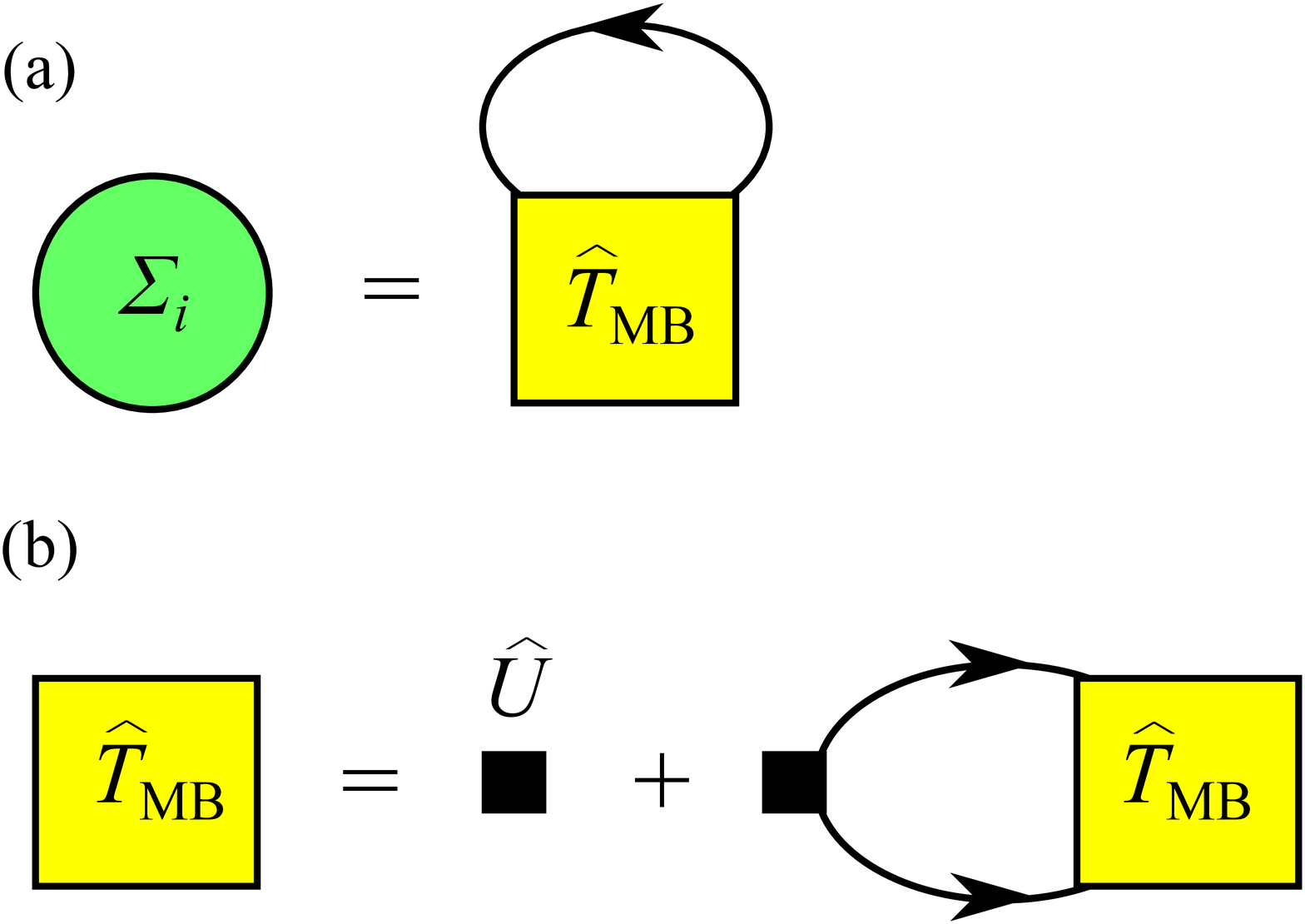}
\caption{Feynman diagrams of (\textbf{a}) the self-energy $\Sigma_i$ and (\textbf{b}) many body $T$ matrix $T_{\rm MB}$ where single lines represent bare Green's functions $G^0_i$.
The black box indicates the matrix of coupling constants $\hat{U}$.
}
\label{fig2}
\end{figure}

We examine the effects of pairing fluctuations on thermodynamic quantities by extending the many body $T$ matrix approach to the present two band system.
We define the Matsubara--Green's function in the $i$ band as:
\begin{eqnarray}
\label{eq3}
G_i(\bm{k},i\omega_{\ell})=\frac{1}{i\omega_{\ell}-\xi_{\bm{k},i}-\Sigma_{i}(\bm{k},i\omega_\ell)},
\end{eqnarray}
where $\omega_\ell=(2\ell+1)\pi T$ ($\ell$ integer) is a fermionic Matsubara frequency at temperature $T$.
As~diagrammatically drawn in Figure \ref{fig2}a, the self-energy $\Sigma_{i}(\bm{k},i\omega_\ell)$ contains pair-fluctuation corrections~as:
\begin{eqnarray}
\label{eq4}
\Sigma_i(\bm{k},i\omega_\ell)=T\sum_{\bm{Q}}\sum_l[\hat{T}_{\rm MB}(\bm{Q},i\nu_l)]_{ii}G_i^0(\bm{k}-\bm{Q},i\omega_\ell-i\nu_l),
\end{eqnarray}
where $G_i^0(\bm{k},i\omega_\ell)=(i\omega_\ell-\xi_{\bm{k},i})^{-1}$ is the bare Green's function, {$\nu_l=2l\pi T$ ($l$ integer) is a bosonic Matsubara frequency,} and:
\begin{eqnarray}
\label{eq5}
\hat{T}_{\rm MB}(\bm{Q},i\nu_l)=[1+\hat{U}\hat{\Pi}(\bm{Q},i\nu_l)]^{-1}\hat{U}
\end{eqnarray}
is the $2 \times 2$ many body $T$ matrix (see also Figure~\ref{fig2}b).
Here, we have defined the matrix of coupling constants as:
\begin{eqnarray}
\label{eq6} 
\hat{U}=
\left(\begin{array}{cc}
U_{11} & U_{12} \\
U_{12} & U_{22} 
\end{array}
\right).
\end{eqnarray}

The particle-particle propagator matrix $\hat{\Pi}(\bm{Q},i\nu_l)$ is given by:
\begin{eqnarray}
\label{eq7}
\hat{\Pi}(\bm{Q},i\nu_l)=
\left(\begin{array}{cc}
\Pi_{11}(\bm{Q},i\nu_l) & 0 \\
0 & \Pi_{22}(\bm{Q},i\nu_l) 
\end{array}
\right),
\end{eqnarray}
with:
\begin{eqnarray}
\Pi_{jj}(\bm{Q},i\nu_l)&=&T\sum_{\bm{k}}^{k_0}\sum_\ell G^{0}_{j}(\bm{k}+\bm{Q},i\omega_\ell+i\nu_l)
G^{0}_{j}(-\bm{k},-i\omega_\ell).\nonumber\\
&=&-\sum_{\bm{k}}^{k_0}\frac{1-f(\xi_{\bm{k+Q}/2,j})-f(\xi_{\bm{-k+Q}/2,j})}{i\nu_l-\xi_{\bm{k+Q}/2,j}-\xi_{-\bm{k+Q}/2,j}},
\label{eq8}
\end{eqnarray}
where $f$ is the Fermi function at temperature $T$. 
\par
{Note that the particle-particle propagator matrix $\hat{\Pi}(\bm{Q},i\nu_l)$ is diagonal because in the Hamiltonian (\ref{eq1}), the~two bands are coupled only by the Josephson interaction $U_{12}$ associated with the transfer of a pair from one band to the other one (as is usually assumed, following \cite{SMW}). Off-diagonal terms would appear only in the presence of cross--pairing interaction directly coupling fermions belonging to different bands. The~self-energy (\ref{eq4}), which is obtained by applying standard Feynman rules for the perturbative expansion of the single particle Green's function, sums up an infinite series of ladder diagrams describing the interaction of a fermion in a given band with a fluctuating pair. Mathematically, the fluctuating pair interacting with fermions belonging to the band $i$ is described by the matrix element $[\hat{T}_{\rm MB}]_{ii}$ of the many body $T$ matrix. Physically, it corresponds to a pair that starts and ends in a given band after having been transferred back and forth from and to the other band (as can be seen by expanding Equation~(\ref{eq1}) in powers~of~$\hat{U}\hat{\Pi}$).} 

In the following, we mention how to calculate the physical quantities of interest {and the practical differences with respect to the NSR approach used in our previous work \cite{Tajima1}.
To determine the chemical potential $\mu$,
we numerically invert the particle number equation $n=n_1+n_2$ with:
\begin{eqnarray}
\label{eq9}
n_i=2\sum_{\bm{k}}\bar{n}_i(k),
\end{eqnarray}
where $\bar{n}_i(k)$ is the momentum distribution function in the $i$ band, which
is obtained from the Matsubara frequency summation of the dressed Green's function $G_i(\bm{k},i\omega_{\ell})$ with $T$ matrix self-energy $\Sigma_i(\bm{k},i\omega_{\ell})$ as:
\begin{eqnarray}
\label{eqadd1}
\bar{n}_i(k)=T\sum_{\ell}G_i(\bm{k},i\omega_{\ell})e^{i\omega_{\ell}0^+}.
\end{eqnarray}

The delicate frequency sum in Equation~(\ref{eqadd1}), with the convergence factor $e^{i\omega_{\ell}0^+}$, is performed by adding and subtracting in the sum a non-interacting Green's function, leading to: 
\begin{eqnarray}
\label{eqadd2}
\bar{n}_i(k)=f(\xi_{\bm{k},i}) + T\sum_{\ell}G_i(\bm{k},i\omega_{\ell})\Sigma_i(\bm{k},i\omega_\ell)G_i^0(\bm{k},i\omega_\ell),
\end{eqnarray}
where now, the sum over the frequency is absolutely convergent, and the convergence factor can be~dropped.}

We note that the difference between the present $T$ matrix approach and the Nozi\`{e}res--Schmitt--Rink one~\cite{NSR,SadeMelo} adopted in our previous work~\cite{Tajima1} consists in the form of $G_i(\bm{k},i\omega_n)$ used in Equation~(\ref{eqadd1}). 
Specifically, in the NSR approach, Dyson's Equation (\ref{eq3}) for $G_i$ is truncated to first order as:
\begin{eqnarray}
G_i^{\rm NSR}(\bm{k},i\omega_\ell)=G_i^0(\bm{k},i\omega_\ell)+G_i^0(\bm{k},i\omega_\ell)\Sigma_i(\bm{k},i\omega_\ell)G_i^0(\bm{k},i\omega_\ell),
\end{eqnarray}
while in the $T$ matrix approach, Equation~(\ref{eq3}) is retained exactly.
{In the NSR approach, the fermions in a given band thus interact only once with a fluctuating pair, while the full solution of Dyson's equation performed in the $T$ matrix approach retains all repeated scattering of independent fluctuations.}
{On~the numerical side, a simplification introduced by the NSR approach is that the NSR particle number \mbox{{$n^{\rm NSR}=n_{1}^{\rm NSR}+n_{2}^{\rm NSR}$}}, with:
\begin{eqnarray}
\label{eqadd3}
n_i^{\rm NSR}=2T\sum_{\bm{k}} \sum_{\ell}G_i^{\rm NSR}(\bm{k},i\omega_\ell)e^{i\omega_{\ell}0^+},
\end{eqnarray}
can also be obtained from the Equation~\cite{Iskin}:
\begin{eqnarray}
\label{eqadd4}
n^{\rm NSR}=2\sum_{\bm{k}}f(\xi_{\bm{k},i})-T\sum_{\bm{Q}}\sum_l\frac{\partial}{\partial \mu}{\rm ln}{\rm det}\left[1+\hat{U}\hat{\Pi}(\bm{Q},i\nu_{l})\right],
\end{eqnarray}
which with a numerical differentiation with respect to $\mu$, avoids the calculation of the self-energy and the nested sums over ${\bm k},l,{\bm Q},\ell$ required by Equations~(\ref{eq4}) and (\ref{eqadd3}). The implementation of these nested sums is instead unavoidable in the $T$ matrix approach.
Clearly, the NSR and $T$ matrix approaches are essentially equivalent when the self-energy corrections are small, while when this does not occur, quantitative differences are expected in their results for thermodynamic quantities, such as, e.g., the chemical potential $\mu$ and critical temperature $T_{\rm c}$.
For the BCS-BEC crossover in a single band, these differences turn out to be rather moderate across the whole BCS-BEC crossover, even when the self-energy is not small~\cite{Strinati}.} We will show below that the same occurs in the two band system considered in this work.

%The single-particle density of states can be obtained from the analytic continuation of $G_i(\bm{k},i\omega_\ell)$ as
%\begin{eqnarray}
%\label{eq11}
%N_i(\omega)=-\frac{1}{\pi}\sum_{\bm{k}}{\rm Im}G_i(\bm{k},i\omega_n\rightarrow \omega+i\delta).
%\end{eqnarray}
%The analytic continuation is numerically done by means of the Pad\'{e} approximation~\cite{Serene} with $\delta=5\times 10^{-3}k_{\rm F,t}^2/(2m)$.

%%%%%%%%%%%%%%%%%%%%%%%%%%%%%%%%%%%%%%%%%%
\section{Results}
\begin{figure}[t]
\centering
\includegraphics[width=10 cm]{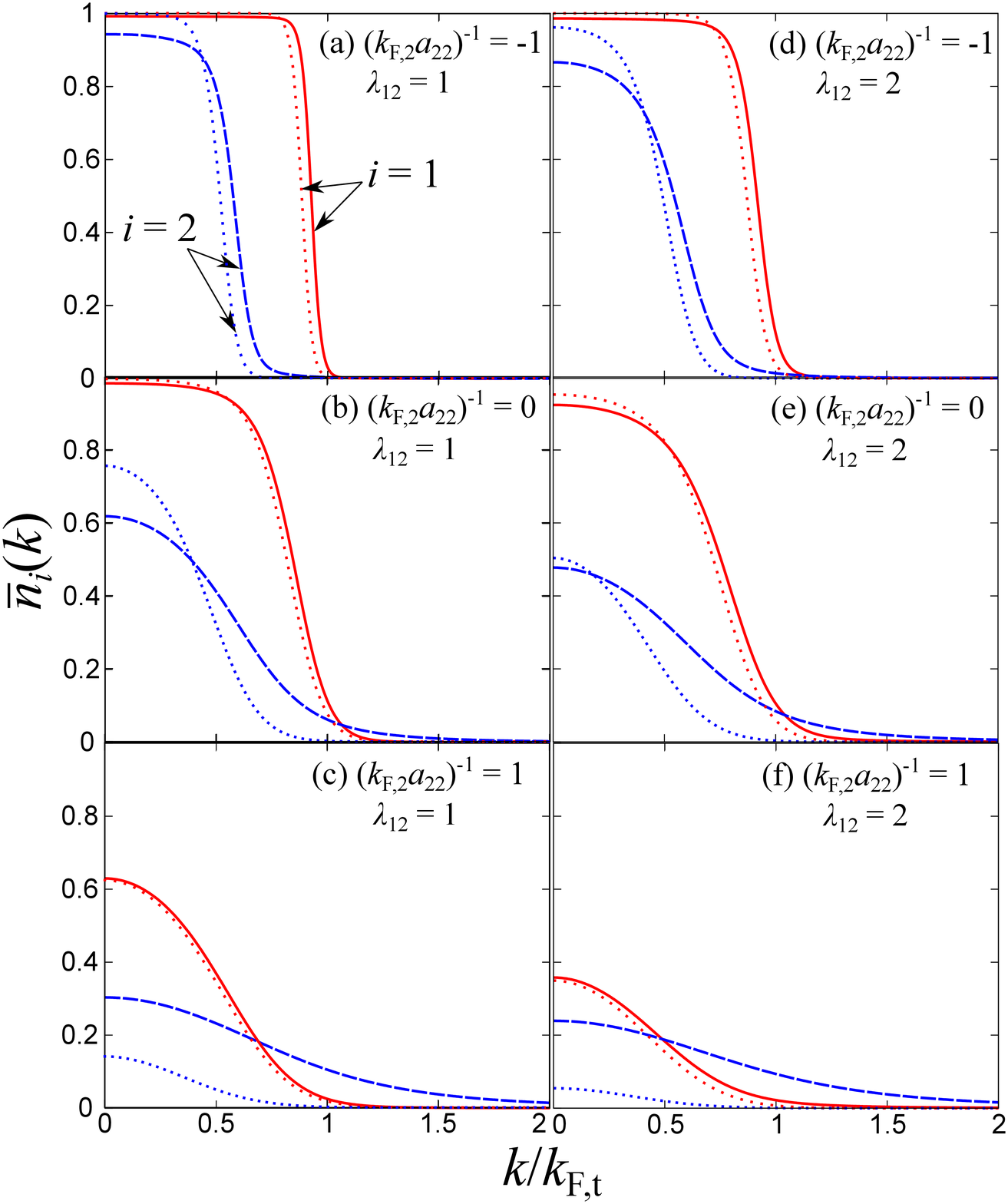}
\caption{{Momentum distribution functions $\bar{n}_i(k)$ at the superfluid critical temperature $T_{\rm c}$ identified by Equation~(\ref{eq10}). The left column corresponds to $\lambda_{12}=1$ and (\textbf{a}) $(k_{\rm F,2}a_{22})^{-1}=-1$, (\textbf{b}) $(k_{\rm F,2}a_{22})^{-1}=0$, and (\textbf{c}) $(k_{\rm F,2}a_{22})^{-1}=1$, while the right column corresponds to $\lambda_{12}=2$ and (\textbf{d}) $(k_{\rm F,2}a_{22})^{-1}=-1$, (\textbf{e})~$(k_{\rm F,2}a_{22})^{-1}=0$, and (\textbf{f}) $(k_{\rm F,2}a_{22})^{-1}=1$. The coupling in the deep band is fixed at $(k_{\rm F,1}a_{11})^{-1}=-4$. The~dotted curves correspond to the free Fermi distribution functions $f(\xi_{\bm{k},i})$ calculated with the same chemical potential and~temperature.}
}
\label{figadd1}
\end{figure}

{As a first quantity, Figure~\ref{figadd1} shows our results for the momentum distribution functions $\bar{n}_i(k)$ in the two bands at three different values of the coupling strength $(k_{\rm F,2}a_{22})^{-1}$ spanning the BCS-BEC crossover. The interband coupling parameter is taken at the values $\lambda_{12}=1$ (left column) and $\lambda_{12}=2$ (right column), while the temperature is set at the superfluid critical temperature $T_c$, as determined by Equation~(\ref{eq10}) (the~results for $T_c$ are presented in Figure \ref{fig3} below).
In the weak coupling cases shown in Figure~\ref{figadd1}a,d, where~$(k_{\rm F,2}a_{22})^{-1}=-1$,
the momentum distribution $\bar{n}_i(k)$ in each band exhibits a clear Fermi step, rounded by finite temperature.
Indeed, when $\Sigma_i(\bm{k},i\omega_{\ell}) \simeq 0$, one sees from Equation~(\ref{eqadd2}) that $\bar{n}_i(k)$ is dominated by the Fermi--Dirac distribution function $f(\xi_{\bm{k},i})$.
At the stronger couplings \mbox{$(k_{\rm F,2}a_{22})^{-1}=0$ and $(k_{\rm F,2}a_{22})^{-1}=1$} shown in Figure~\ref{figadd1}b,c,e,f, respectively, the interaction effects become progressively more important. In~the shallow band, in particular, the momentum distribution clearly departs from the corresponding non-interacting distributions for both values of the pair-exchange coupling. 
For~the deep band, instead, the effects are much weaker. This is because increasing the interaction in the shallow band affects the fermions in the deep band only indirectly, through the pair-exchange coupling $\lambda_{12}$. One~sees indeed that the interaction effects on $\bar{n}_1(k)$ are more visible for the larger value of the pair-exchange~coupling.}

Figure \ref{fig3} shows the calculated superfluid critical temperature $T_{\rm c}$ for various values of the pair-exchange coupling $\lambda_{12}$ as functions of the coupling $(k_{\rm F,2}a_{22})^{-1}$ in the shallow band.
{The superfluid critical temperature is determined by employing the Thouless criterion, which for the present multi-band system, corresponds to the equation: 
\begin{eqnarray}
\label{eq10}
{\rm det}[1+\hat{U}\hat{\Pi}(\bm{Q}=0,i\nu_l=0)]=0.
\end{eqnarray}

\begin{figure}[t]
\centering
\includegraphics[width=8 cm]{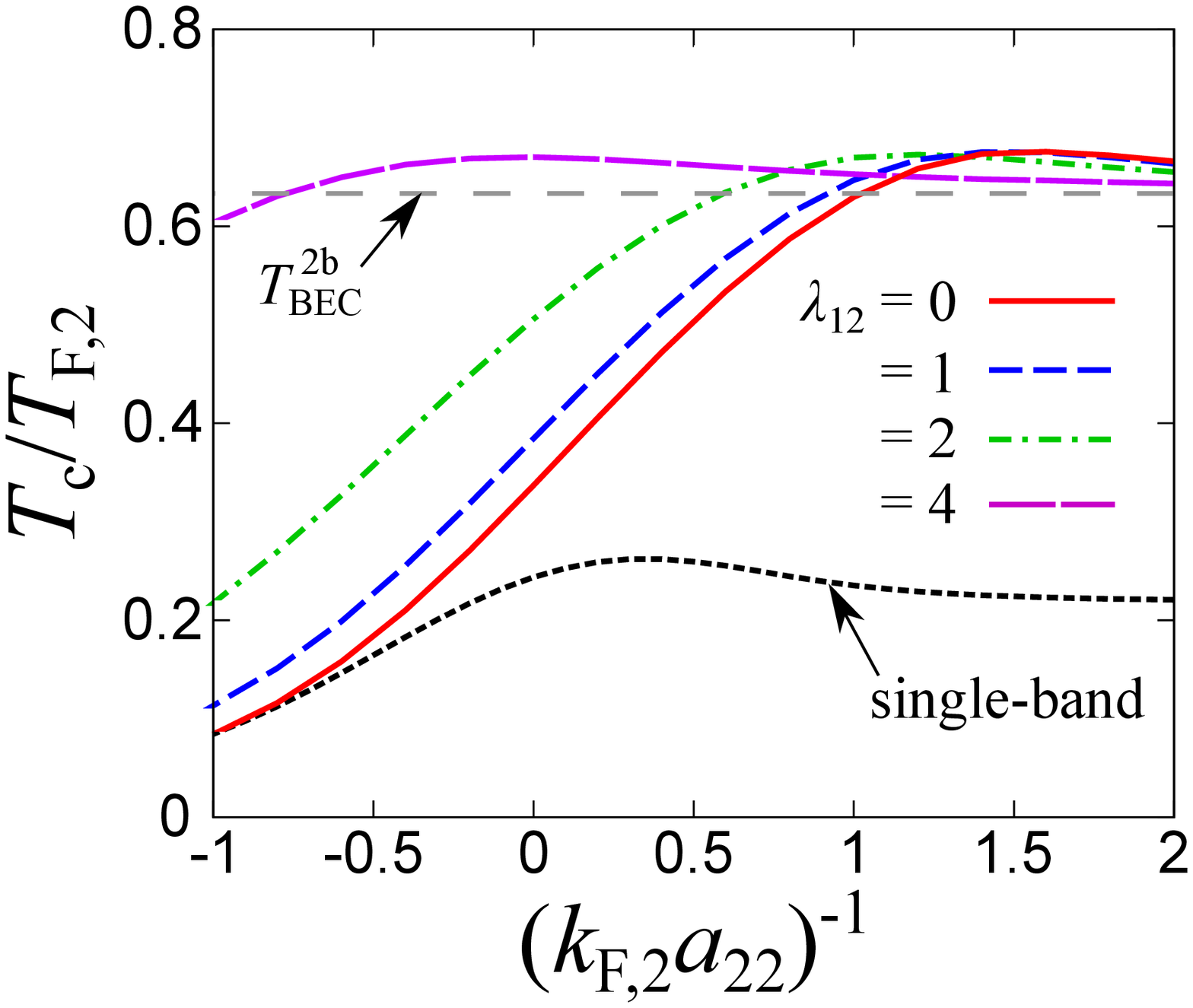}
\caption{Superfluid critical temperature $T_{\rm c}$ in the Bardeen--Cooper--Schrieffer--Bose--Einstein condensation (BCS-BEC) crossover regime of the shallow band calculated for different values of the pair-exchange coupling $\lambda_{12}$ as a function of the coupling $(k_{\rm F,2}a_{22})^{-1}$ in the shallow band.
The black dotted curve represents the numerical result of the single band counterpart.
The dashed line $T_{\rm BEC}^{\rm 2b}=0.218(n/n^0_2)^{2/3}T_{\rm F,2}=0.633T_{\rm F,2}$ is the molecular BEC temperature when all particles in both bands form tightly bound molecules. {The~coupling in the deep band is fixed at $(k_{\rm F,1}a_{11})^{-1}=-4$.}
}
\label{fig3}
\end{figure} 

Physically, the Thouless criterion corresponds to a diverging pairing susceptibility in any of the two bands. Coming from the normal phase, it can be obtained by calculating the pairing susceptibility within linear response theory, with the many body $T$ matrix diagrams as the central part of the corresponding two particle Green's function (see, e.g.,~\cite{Pekker}). Coming instead from the superfluid phase, the Thouless criterion corresponds to the vanishing of the BCS gap in both bands. We stress in this respect that, as for the single band case, the differences in the value of the critical temperature compared with the BCS mean field are entirely due to the different particle number equation, which includes the effects of paring fluctuations in the $T$ matrix approach.} 
One can see that $T_{\rm c}$ in the two band system is greatly enhanced in the strong coupling regime for all values of $\lambda_{12}$, compared to the isolated single band counterpart.
For $\lambda_{12}=0$, this behavior originates from the particle transfer from the deep band to the shallow band, as a result of free energy minimization.
Indeed, in the strong coupling limit, $T_{\rm c}$ approaches the BEC temperature $T_{\rm BEC}^{\rm 2b}=0.218(n/n_2^0)^{\frac{2}{3}}T_{\rm F,2}=0.633T_{\rm F,2}$, where all particles in both bands form tightly bound molecules.
Additionally, if one increases $\lambda_{12}$, the BEC limit is reached already at rather weak values of the coupling parameter $(k_{\rm F,2}a_{22})^{-1}$.
This finding can be understood as the increase of an effective intraband coupling due to the pair-exchange induced interaction as:
\begin{eqnarray}
\label{eq12}
U_{\rm 22}^{\rm eff}(\bm{Q},i\nu_l)=U_{22}-\frac{U_{12}U_{21}\Pi_{11}(\bm{Q},i\nu_l)}{1+U_{11}\Pi_{22}(\bm{Q},i\nu_l)}.
\end{eqnarray}

From Equation (\ref{eq12}), one can see that the two body attractive interaction in the shallow band is enhanced due to the pair-exchange coupling. 
\par
\begin{figure}[t]
\centering
\includegraphics[width=8 cm]{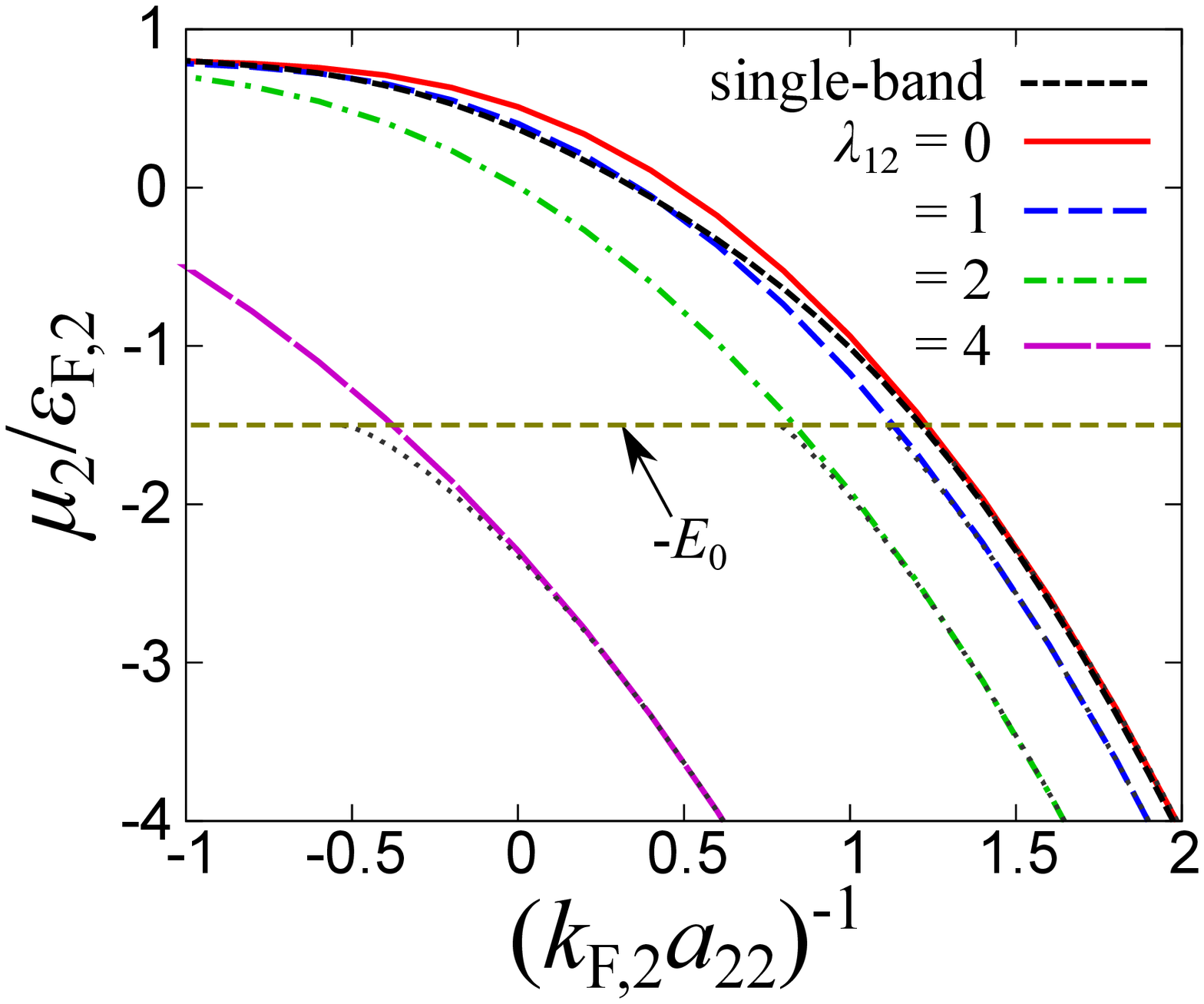}
\caption{Chemical potential $\mu_2\equiv\mu-E_0$ referred to the bottom of the shallow band, at $T=T_{\rm c}$ for different values of the pair-exchange coupling $\lambda_{12}$ as a function of the coupling $(k_{\rm F,2}a_{22})^{-1}$ in the shallow band.
The~horizontal line $-E_0$ corresponds to the bottom of the lower band. 
The~dotted curves represent half of the two body binding energy $-E_{\rm b}/2$ in our two band configuration.{ The coupling in the deep band is fixed at $(k_{\rm F,1}a_{11})^{-1}=-4$.}}
\label{fig4}
\end{figure}

The effect of the pair-exchange coupling induced intraband attraction can also be found in the evolution of the chemical potential at $T_c$. Specifically, for the sake of 
comparison with the single band case, it is useful to consider the chemical potential $\mu_2\equiv\mu-E_0$ referred to the bottom of the shallow band, as~presented in 
Figure \ref{fig4}.
 One sees that by increasing the pair-exchange coupling, the BEC limit for the shallow band (corresponding roughly to $\mu_2 < 0$) is reached very rapidly 
compared with the single-band case. 
When the chemical potential $\mu$ goes further below and crosses the bottom of the lower band (corresponding to the dashed horizontal line 
$\mu_2=-E_0$ in Figure \ref{fig4}), the calculated values of the chemical potential can be compared 
with the bound state solution of the two body Schr\"{o}dinger equation:
%\begin{eqnarray}
%\label{eq13}
%\frac{k^2}{m}\psi_1(\bm{k})+
%\sum_{\bm{k}'}\left[U_{11}\psi_1(\bm{k}')+U_{12}\psi_2(\bm{k}')\right]=E\psi_1(\bm{k}),
%\end{eqnarray}
%\begin{eqnarray}
%\label{eq14}
%\left(\frac{k^2}{m}+2E_{0}\right)\psi_2(\bm{k})+
%\sum_{\bm{k}'}\left[U_{22}\psi_2(\bm{k}')+U_{21}\psi_2(\bm{k}')\right]=E\psi_2(\bm{k}),
%\end{eqnarray}
\begin{equation}
\begin{cases}
\frac{k^2}{m}\psi_1(\bm{k})+
\sum_{\bm{k}'}\left[U_{11}\psi_1(\bm{k}')+U_{12}\psi_2(\bm{k}')\right]=E\psi_1(\bm{k})&\\
\left(\frac{k^2}{m}+2E_{0}\right)\psi_2(\bm{k})+
\sum_{\bm{k}'}\left[U_{22}\psi_2(\bm{k}')+U_{21}\psi_2(\bm{k}')\right]=E\psi_2(\bm{k}),&
\end{cases}
\end{equation}
where $\psi_i$ is the two body wave function in the $i$ band with $E= 2E_0-E_{\rm b}$ 
(and the binding energy $E_{\rm b}$ is referred to the bottom of the lower band).
After a straightforward calculation, we obtain the equation for the binding energy $E_{\rm b}$ as:
%\begin{eqnarray}
%\label{eq15}
%&&\left[1+\frac{mk_0U_{11}}{2\pi^2}\left\{1-\frac{\sqrt{m|E_1|}}{k_0}\tan^{-1}\left(\frac{k_0}{\sqrt{m|E_1|}}\right)\right\}\right]\cr
%&\times&\left[1+\frac{mk_0U_{22}}{2\pi^2}\left\{1-\frac{\sqrt{m|E_2|}}{k_0}\tan^{-1}\left(\frac{k_0}{\sqrt{m|E_2|}}\right)\right\}\right]\cr
%&-&\left(\frac{mk_0U_{12}}{2\pi^2}\right)^2\left\{1-\frac{\sqrt{m|E_1|}}{k_0}\tan^{-1}\left(\frac{k_0}{\sqrt{m|E_1|}}\right)\right\}\cr
%&\times&\left\{1-\frac{\sqrt{m|E_2|}}{k_0}\tan^{-1}\left(\frac{k_0}{\sqrt{m|E_2|}}\right)\right\}=0,
%\end{eqnarray}
\begin{eqnarray}
\label{eq15}
&&\prod_{i=1,2}\left\{1+\frac{mk_0U_{ii}}{2\pi^2}\left[1-\frac{\sqrt{m|E_i|}}{k_0}\tan^{-1}\left(\frac{k_0}{\sqrt{m|E_i|}}\right)\right]\right\}\cr
&&-\left(\frac{mk_0U_{12}}{2\pi^2}\right)^2\prod_{i=1,2}\left[1-\frac{\sqrt{m|E_i|}}{k_0}\tan^{-1}\left(\frac{k_0}{\sqrt{m|E_i|}}\right)\right]=0,
\end{eqnarray}
where $E_{i}=-E_{\rm b}+2E_0\delta_{i,1}$.
In Figure \ref{fig4}, one can find that $\mu_2$ is indeed in good agreement with $-E_{\rm b}/2$ in the strong coupling regime.
%$E_{\rm b} $ increases with increasing $\lambda_{12}$ due to the crossband induced attraction as given by Equation (\ref{eq12}).
%For comparison, we also plot the results of single-band counterpart.
%The two band result coincides with single-band one in the case of vanishing interband coupling.

\par
\begin{figure}[t]
\centering
\includegraphics[width=8 cm]{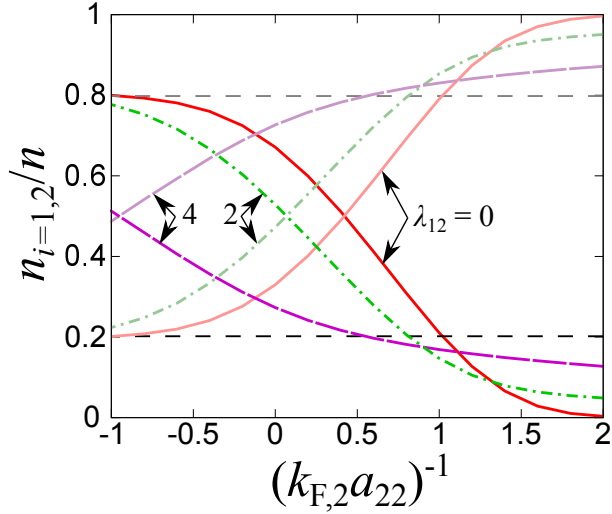}
\caption{Occupation number density $n_1$ (decreasing functions) and $n_2$ (increasing functions) at $T=T_{\rm c}$ for different values of the pair-exchange coupling $\lambda_{12}$ as a function of the coupling $(k_{\rm F,2}a_{22})^{-1}$ in the shallow band.
The horizontal dashed lines represent $n_{1,0}=0.798n$ (upper line) and $n_{2,0}=0.202n$ (lower line). {The~coupling in the deep band is fixed at $(k_{\rm F,1}a_{11})^{-1}=-4$.}
}
\label{fig5}
\end{figure}

To confirm our scenario for the enhancement of $T_{\rm c}$ due to the particle transfer between bands,
we~also calculate the occupation number density $n_i$ in each band.
Figure \ref{fig5} shows the calculated number densities $n_{i=1,2}$ in the BCS-BEC crossover regime.
In the weak coupling limit, $n_{1}$ and $n_{2}$ approach the non-interacting results $n_{1,0}=0.798n$ and $n_{2,0}=0.202n$, as expected.
For increasing coupling $(k_{\rm F,2}a_{22})^{-1}$,
$n_{2}$ increases because the particles in the deep band flow into the shallow band.
Note that a full transfer of particles from the deep to the shallow band can be achieved even for vanishing pair-exchange coupling for sufficiently strong interaction in the shallow band. On the other hand, a finite pair-exchange coupling induces the mixing of two kinds of molecular BEC in the strong coupling regime~\cite{Tajima1}.
As a result, $n_{1}$ remains finite in the strong coupling limit with finite $\lambda_{12}$. 

\begin{figure}[t]
\centering
\includegraphics[width=8 cm]{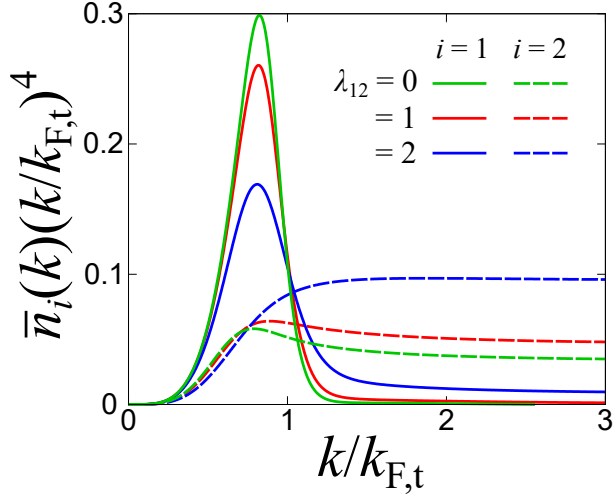}
\caption{Momentum distribution functions $\bar{n}_i(k)$ multiplied by $(k/k_{\rm F,t})^4$ at $(k_{\rm F,2}a_{22})^{-1}=0$ and $T=T_{\rm c}$.
Solid (dashed) lines show the results of a deep (shallow) band. {The coupling in the deep band is fixed at~$(k_{\rm F,1}a_{11})^{-1}=-4$.}
}
\label{figadd2}
\end{figure}

{The $T$ matrix approach presented in this paper can address the pair correlations from the momentum distribution functions $\bar{n}_i(k)$.
Figure~\ref{figadd2} shows the calculated $\bar{n}_i(k)$ multiplied by $(k/k_{\rm F,t})^4$ at $(k_{\rm F,2}a_{22})^{-1}=0$, $T=T_{\rm c}$ (and $(k_{\rm F,1}a_{11})^{-1}=-4$, as everywhere in our paper).
In the single band case, the momentum distribution function has a large momentum tail characterized by Tan's contact $C$~\cite{Tan1,Tan2,Tan3}. 
Since Tan's contact is defined as a thermodynamic quantity
conjugate to the inverse scattering length,
the two band system in principle involves two contact parameters $C_i$ as:
\begin{eqnarray}
\label{eqC}
\bar{n}_i(k)\xrightarrow[k\rightarrow\infty]{}\frac{C_i}{k^4}.
\end{eqnarray}

Indeed, one can find that $\bar{n}_i(k)k^4$ in both bands approach constants (namely, $C_i$) in the large momentum region as in the case of single band~\cite{PalestiniC}.
While $C_1$ is small because of the weak intraband coupling $(k_{\rm F,1}a_{11})^{-1}=-4$, $C_2$ in the strongly interacting shallow band is larger. 
Moreover, at finite $\lambda_{12}$, both $C_1$ and $C_2$ increase due to the pair-exchange induced interaction given by Equation~(\ref{eq12}).
Since $C_{i}$ is proportional to the molecular density in the strong coupling limit, the emergence of two finite contact parameters supports the coexistence of two kinds of molecular BEC in the present two band system.}

\begin{figure}[t]
\centering
\includegraphics[width=8 cm]{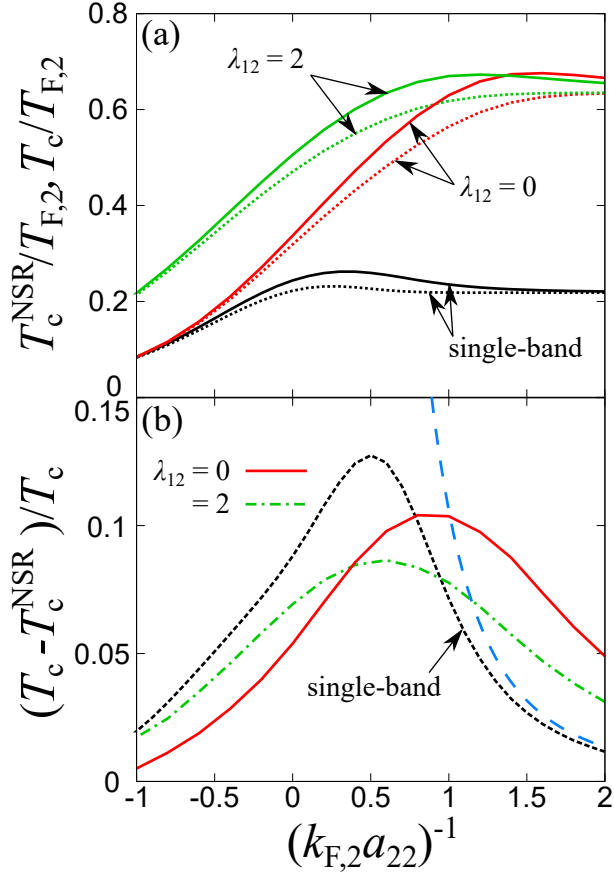}
\caption{(\textbf{a}) Critical temperatures within the $T$ matrix approach ($T_{\rm c}$, full lines) and Nozi\`eres--Schmitt--Rink (NSR) approximation ($T_{\rm c}^{\rm NSR}$, dotted lines) for different values of the coupling $\lambda_{12}$ as a function of the coupling $(k_{\rm F,2}a_{22})^{-1}$ in the shallow band.
(\textbf{b}) Relative difference between the two critical temperatures $(T_{\rm c}-T_{\rm c}^{\rm NSR})/T_{\rm c}$ along the BCS-BEC crossover in the shallow band.
The dashed curve is the asymptotic behavior for the single band counterpart given by Equation (\ref{eq17})~\cite{Pini}. {The coupling in the deep band is fixed at $(k_{\rm F,1}a_{11})^{-1}=-4$.}}
\label{fig6}
\end{figure}

The above results for the critical temperature, chemical potential, and particle numbers obtained with the $T$ matrix approach thus confirm our previous findings obtained with the NSR approximation~\cite{Tajima1}.
{We~emphasize, however, that our results presented in this paper are different from those of NSR.}
To make this comparison more quantitative, we report in Figure \ref{fig6}a the critical temperature obtained by the $T$ matrix approach ($T_{\rm c}$) and by the NSR approximation ($T_{\rm c}^{\rm NSR}$).
One can see that, also in the two band case, both~approaches exhibit a qualitatively similar behavior along the BCS-BEC crossover.
Indeed, as shown in Figure \ref{fig6}b, the relative difference between the critical temperature obtained by the two approaches, namely $(T_{\rm c}-T_{\rm c}^{\rm NSR})/T_{\rm c}$, is less than $15\%$ in the whole crossover region. An interesting feature of Figure~\ref{fig6}b is also the position of the maximum of $(T_{\rm c}-T_{\rm c}^{\rm NSR})/T_{\rm c}$. Since the expansion parameter of Dyson's equation is $G^0_i\Sigma_i$, one expects the maximum to occur for intermediate couplings. This is because in weak coupling, $\Sigma_i$~is small due to weak interaction, while for strong coupling, it is the large and negative chemical potential contained in $G^0_i$ that makes the expansion parameter small. This expectation is confirmed by Figure~\ref{fig6}b. In particular, for a single band, the maximum is located near the coupling strength where the chemical potential goes below the bottom of the single band (corresponding to $\mu_2=0$ in Figure \ref{fig4}). The same criterion holds also in the two band case, where now, the relevant band is the deep one (corresponding to $\mu=0$ and thus $\mu_2=-E_0$ in Figure \ref{fig4}). 

Notice furthermore that, as one can evinces from Figure \ref{fig6}a, in the strong coupling limit, the $T$ matrix approach reaches the BEC value for $T_c$ more slowly than the NSR approach. 
In the single band case, one~can show indeed that for the $T$ matrix approach, the BEC value for $T_c$ is reached in the strong coupling limit according to the asymptotic law~\cite{Pini}:
\begin{eqnarray}
\label{eq17}
\frac{T_{\rm c}-T_{\rm BEC}}{T_{\rm BEC}}=\frac{(k_{\rm F}a)^3}{3\pi},
\end{eqnarray}
where all quantities refer to a single band, for which $T_{\rm BEC}=0.218T_{\rm F}$. A similar power law behavior is expected in the two band case, with a coefficient depending on $\lambda_{12}$. For the NSR approximation, on the other hand, one can easily show that the correction to $T_{\rm BEC}$ vanishes exponentially in the strong coupling limit (for both the single band and two band cases). For the single band case, Equation (\ref{eq17}) shows indeed good agreement with $(T_{\rm c}-T_{\rm NSR})/T_{\rm c}$ in the strong coupling regime (see the dashed line in Figure \ref{fig6}b).

\section{Conclusions}
We investigated the BCS-BEC crossover and the effects of pairing fluctuations for a two band system within the framework of the many body $T$ matrix approximation.
We analyzed the evolution of thermodynamic properties from BCS to BEC limits in a shallow band coupled with a weakly interacting deep band.
We confirmed the finding, previously obtained with the simpler NSR approximation, of a strong enhancement of the critical temperature in the two band case compared with the single band case. 
{Furthermore, we calculated the momentum distribution functions in each band, which exhibit a large momentum tail characterized by Tan's contact.
Generalizing Tan's relation to the present two band model is an interesting future problem.}
We finally remark that the calculation of dynamic quantities (such as single particle spectral weight functions and density of states) should necessarily rely on approaches like the $T$ matrix approximation, which do not expand Dyson's equation. {For such quantities, therefore, the~NSR approach could not be used}. 
Our analysis of thermodynamic quantities with the $T$ matrix approach thus paves the way toward future investigations of spectral weight functions of a two band system through the BCS-BEC crossover.

%%%%%%%%%%%%%%%%%%%%%%%%%%%%%%%%%%%%%%%%%%
%% optional
%\supplementary{The following are available online at \linksupplementary{s1}, Figure S1: title, Table S1: title, Video S1: title.}

% Only for the journal Methods and Protocols:
% If you wish to submit a video article, please do so with any other supplementary material.
% \supplementary{The following are available at \linksupplementary{s1}, Figure S1: title, Table S1: title, Video S1: title. A supporting video article is available at doi: link.}

%%%%%%%%%%%%%%%%%%%%%%%%%%%%%%%%%%%%%%%%%%
%\authorcontributions{Conceptualization, H.T., A.P., and P.P.; methodology, H.T., A.P., and P.P., software, H.T.; investigation, H.T.; data curation, H.T.; writing -- original draft, H.T.; writing -- review and editing, H.T., A.P., and P.P.; visualization, H.T.; supervision, A.P. and P.P.; project administration, A.P. and P.P.}

%%%%%%%%%%%%%%%%%%%%%%%%%%%%%%%%%%%%%%%%%%
%\funding{This research was funded by the Japan Society for Promotion of Science with Grant-in-Aid for JSPS fellows No.~17J03975 and grant for Scientific Research No.~18H05406, and by Italian MIUR through the PRIN 2015 program, grant No.~2015C5SEJJ001.
%}

%%%%%%%%%%%%%%%%%%%%%%%%%%%%%%%%%%%%%%%%%%
\acknowledgments{We thank Yuriy Yerin for useful discussions. H.T thanks the Physics Division at the University of Camerino for the hospitality.
This research was funded by the Japan Society for Promotion of Science with Grant-in-Aid for JSPS fellows No.~17J03975 and grant for Scientific Research No.~18H05406, and by Italian MIUR through the PRIN 2015 program, grant No.~2015C5SEJJ001.
}

\end{document}